\documentclass[11pt]{article}

 \usepackage{amsmath}
 \usepackage{graphicx}
 \usepackage{amssymb}
 \usepackage{amsfonts}
 \usepackage{latexsym}

\renewcommand{\baselinestretch}{1.1}
\def\beq{\begin{eqnarray}}
\def\eeq{\end{eqnarray}}
\def\ln{\,\mbox{ln}\,}
\def\Ln{\,\mbox{Ln}\,}
\def\Det{\,\mbox{Det}\,}
\def\det{\,\mbox{det}\,}
\def\tr{\,\mbox{tr}\,}

\def\al{\alpha}
\def\be{\beta}

\def\ga{\gamma}
\def\de{\delta}

\def\la{\lambda}
\def\na{\nabla}
\def\pa{\partial}

\def\si{\sigma}

\def\ph{\varphi}

\def\Ga{\Gamma}
\def\De{\Delta}

\begin{document}

\begin{center}

{\Large\bf Universality and ambiguity in fermionic effective actions}
\vskip 6mm


\textbf{Guilherme de Berredo-Peixoto\footnote{
E-mail: guilherme@fisica.ufjf.br.},
\
Dante D. Pereira\footnote{
E-mail: dante.pereira@fisica.ufjf.br.}
\ and \
Ilya L. Shapiro \footnote{
E-mail: shapiro@fisica.ufjf.br. On leave from Tomsk State
Pedagogical University, Tomsk, Russia.}}
\vskip 4mm

Departamento de F\'{\i}sica - ICE,
Universidade Federal de Juiz de Fora, 36036-330, MG, Brazil
\end{center}

\vskip 12mm

\begin{quotation}
\noindent
{\large\bf Abstract}.
\quad
We discuss an ambiguity in the one-loop effective action
of massive fields which takes place in massive fermionic
theories. The universality of logarithmic UV divergences in
different space-time dimensions leads to the non-universality
of the finite part of effective action, which can be called the non-local
multiplicative anomaly. The general criteria of existence of
this phenomena are formulated and applied to fermionic operators
with different external fields.

\end{quotation}
\vskip 4mm

\section{Introduction}

The Effective Action (EA) formalism is an important element of the 
modern Quantum Field Theory (QFT). The consistent use of this 
formalism enables one to deal with very general kind of QFT problems 
and in some cases to go beyond the traditional $S$-matrix approach. 
This is especially important in case of gravitational interactions,
where the EA is our main source of information about quantum effects. 
The main two aspects of using the EA approach are to derive it for a 
given QFT system and to take care about its ambiguities. The last 
part is quite relevant, because one has to distinguish real physical 
effects from the apparent properties which depend on the details of 
the calculational technique.

The most well-known ambiguities in QFT are the dependence on the
renormalization point (e.g., on the parameter $\mu$ in the Minimal
Subtraction scheme of renormalization) and gauge-fixing dependence
in gauge theories. Usually the last issue is eliminated on-shell,
but this procedure can be rather non-trivial, especially beyond the 
one-loop approximation. More general, the result strongly depends 
on the renormalization scheme. For example, the renormalization 
group $\be$-functions in massive theories are quite different if 
they are calculated within the simplified Minimal Subtraction (MS) 
scheme or in the more physical momentum-subtraction schemes. At 
low energies the use of the last method enables one to observe 
the decoupling phenomenon, in QED it is the Appelquist and 
Carazzone theorem \cite{AC}.

On the top of those mentioned above, there may be other ambiguities
in the quantum contributions, including the ones we are going to
discuss here. Despite the UV divergences are sometimes regarded
as a main challenge of QFT, there is one curious thing about them,
which will be important for our consideration. Indeed, the leading
logarithmic divergences define the most stable and universal part
of quantum corrections. For example, these divergences are behind
the UV limit of the $\be$-functions, which does not depend on the
renormalization scheme. At one-loop order the EA of average fields 
is in many cases proportional to the
expressions $\,\Ln \Det {\hat H}$, where ${\hat H}$ is some
differential operator depending on these fields. Many manipulations
with such expressions are justified for the UV part, which is related
to the logarithmic divergences, but they may be not valid at all for
the finite non-leading part of the EA. The reason for this special
feature of the UV divergences is as follows. The divergences are
related to the leading logarithmic behavior of the EA (or amplitudes)
and therefore they are always related to the simple logarithmic form
factors, which do not actually depend on the mass of the field
\cite{Salam-51}. On the other hand, the counterterms which are 
necessary to remove the UV divergences are local, and hence one 
can completely control the algebraic structure of the UV divergences 
by looking at the form of the possible local terms in the classical 
action of the theory.

At the same time, the subleading terms are typically nonlocal
and have, in the quantum theory of massive fields, much more
complicated structure. For this reason we may expect them to
be essentially more ambiguous too. It is interesting that, up
to our knowledge, after Salam \cite{Salam-51} nobody explored
the limits of universality of the UV divergences at the formal
level. What we will show here is that the universality of UV
divergences is directly related to the non-universality of the
finite contributions in the massive theories. This phenomena
can be observed in the fermionic determinants and can be called
nonlocal multiplicative anomaly. In what follows we will discuss
this phenomenon for a general fermionic determinant and also
consider in full details the case of a Dirac fermion coupled
to external scalar field by Yukawa interaction.

The paper is organized as follows. In Sect. 2 we present some
general arguments concerning the ambiguous feature of the finite
parts of functional fermionic determinants in the form of
nonlocal multiplicative anomaly. A few particular cases are
briefly addressed in Sect. 3 and in Sect. 4 we present a full
illustrative analysis for the simplest case of a single scalar
background field. In Sect. 5 we draw our conclusions and discuss
the ambiguity due to the nonlocal multiplicative anomaly and
also another one which is local and mass-independent.

\section{General considerations}

Consider the one-loop EA of the Dirac fermion
coupled to some external field. For the sake of generality,
we will deal also with a curved space-time background. The
one-loop EA can be defined via the path integral
\beq
e^{i \bar{\Gamma}^{(1)}}\,=\,
\int D\psi D {\bar \psi}\,\, e^{i S_f}\,,
\label{EA}
\eeq
where the free fermionic action is defined as
\beq
S_f &=& \int d^4x \sqrt{-g}\,\,{\bar \psi} \hat{H} \psi\,,
\nonumber
\\
\hat{H} &=& i\,(\ga^{\mu}\na_{\mu} - im\hat{1} - i\hat{\phi})\,.
\label{act}
\eeq
The $\hat{\phi}$ is a condensed notation for a generic external
field. For example, we can set
\beq
\hat{\phi} = h\ph + h^*\ga^5 \chi + e\ga^\mu A_\mu
+ \eta\ga^5\ga^\mu S_\mu\,+\,...\,,
\label{gen}
\eeq
where $h$ and $h^*$ are Yukawa couplings corresponding to scalar
and axial scalar fields, $e$ is electromagnetic charge, $\eta$ is
nonminimal coupling to the axial vector related to torsion etc.
It is assumed that the one-loop EA consists from the classical 
action of background fields and the one-loop correction which 
will be the subject of our interest,
\beq
{\bar \Ga}^{(1)} \,=\,-\,i\,\Ln \Det \hat{H}\,,
\label{Det}
\eeq
where the $\Det$ does not take into account Grassmann parity.

We will define the expression (\ref{Det}) through the heat-kernel
method and the Schwinger-DeWitt technique, and this requires
reducing the problem to the derivation of $\,\Ln \Det\,\hat{\cal O}$,
where
\beq
\hat{{\cal O}} \,=\,
{\widehat{\Box}} + 2{\hat h}^\mu\na_\mu + {\hat \Pi}\,.
\label{O}
\eeq
In order to make the reduction, one has to multiply
$\hat{H}$ by an appropriate conjugate operator $\hat{H}^*$,
\beq
\hat{{\cal O}} \,=\, \hat{H}\cdot\hat{H}^{*}
\label{conj}
\eeq
and use the relation
\beq
\Ln \Det\,\hat{H} \,=\,\Ln \Det\,\hat{{\cal O}}
\,-\,\Ln \Det\,\hat{H}^*\,.
\label{rel}
\eeq
Indeed, there is more than one option for choosing the
conjugate operator which enables one to use the relation
(\ref{rel}) in an efficient way. The simplest choice is
\beq
\hat{H}^{*}_1\equiv \hat{H}
=i(\ga^{\mu}\na_{\mu} - im\hat{1} - i\hat{\phi})
\label{H1}
\eeq
and therefore
\beq
\Ln \Det\,\hat{H}
\,=\,\frac12\,\Ln \Det\,\big(\hat{H}\hat{H}_1^* \big)\,.
\label{opt 1}
\eeq
An alternative choice of the conjugate operator is
\beq
\hat{H}^{*}_2 =\,i\,\big(\ga^\mu \na_\mu - im\hat{1}\big)\,.
\label{opt 2}
\eeq
This operator does not depend on $\hat{\phi}$ and hence
\beq
\Ln\Det \hat{H}\Big|_{\hat{\phi}}
\,=\, \Ln\Det \big(\hat{H}\hat{H}_2^{*}\big)\Big|_{\hat{\phi}}\,,
\label{opt 21}
\eeq
where the index $\hat{\phi}$ means we are interested 
only in the $\hat{\phi}$-dependent part of EA. It is easy to note 
that if the relation
\beq
\Det({\hat A}\cdot {\hat B})
= \Det{\hat A}\cdot\Det{\hat B}
\label{com}
\eeq
holds for the fermionic functional determinants,
we are going to meet the two equal expressions,
\beq
\frac{1}{2}\,\Ln\Det \big(\hat{H}\hat{H}_1^{*}\big)
\Big|_{\hat{\phi}}
\,=\,\Ln\Det \big(\hat{H}\hat{H}_2^{*}\big)
\Big|_{\hat{\phi}}\,.
\label{crite}
\eeq

As we shall see below, in reality the Eq. (\ref{crite}) is
satisfied for divergencies, but not for the nonlocal finite
parts of the two effective actions. This is nothing else, but
the non-local version of Multiplicative Anomaly (MA)
\cite{MA-0,MA-1,MA-2,MA-3,MA-4}. The possibility of this
mathematical feature of the functional determinants has been
discussed for the long time on the basis of $\zeta$-regularization
(see, e.g., \cite{zeta}). The direct calculations on the
constant curvature background confirmed the existence of MA
\cite{MA-0,MA-1}, but it was soon realized that the difference
can be just a manifestation of the different choice of $\mu$
for the distinct determinants \cite{MA-2,MA-3,MA-4}. The
only safe way to obtain MA is to detect it in the non-local
part of EA, which is qualitatively different from the local
one related to divergences\footnote{Another possibility is 
to consider some unusual version of QFT, e.g., in the presence 
of chemical potential. In this case one can observe a MA in 
the local sector which depends on this parameter and does not 
necessary reduce to the $\mu$-dependence \cite{wipf}.}. 
In this case we will see that the
MA is some new ambiguity of EA and not a particular case of
the well-known $\mu$-dependence.

Before starting practical calculations, let us make some
general observations on the relation (\ref{crite}) for
divergencies and for the finite part of EA. Within the
heat-kernel method, the one-loop EA is given by the
expression (see, e.g., \cite{BDW-65})
\beq
{\bar \Ga}^{(1)}\,=\,i\,
\mbox{sTr}\, \lim\limits_{x^\prime \to x}\,\int\limits_0^\infty
\,\frac{ds}{s}\,\,{\hat U}(x,x^\prime\,;s)\,,
\label{SchDW-1}
\eeq
where the evolution operator satisfies the equation
\beq
i\,\frac{\pa {\hat U(x,x^\prime\,;s)}}{\pa s}
\,=\,-\,\hat{{\cal O}} \,{\hat U(x,x^\prime\,;s)}\,,
\quad
U(x,x^\prime\,;0)=\de(x,x^\prime)\,.
\label{SchDW-2}
\eeq
A useful representation for the evolution operator is
\beq
{\hat U}(x,x^\prime\,;s)\,=\,{\hat U}_0(x,x^\prime\,;s)
\,\sum_{k=0}^{\infty}\,(is)^k\,{\hat a}_k(x,x^\prime)\,,
\label{SchDW-3}
\eeq
where ${\hat a}_k(x,x^\prime)$ are the so-called
Schwinger-DeWitt coefficients,
\beq
{\hat U}_0(x,x^\prime\,;s)
= \frac{{\cal D}^{1/2}(x,x^\prime)}{(4\pi i \,s)^{n/2}}
\,
\exp\Big\{{\frac{i\si (x,x^\prime)}{2s} - m^2s}\Big\},
\label{SchDW-4}
\eeq
$\si (x,x^\prime)$ is the geodesic distance between $x$
and $x^\prime$ points and
${\cal D}$ is the Van Vleck-Morette determinant
\beq
{\cal D}(x,x^\prime)\,=\,\det\left[
-\,\frac{\pa^2 \si (x,x^\prime)}{\pa x^\mu\,\pa x^{\prime\nu}}
\right]\,.
\label{SchDW-5}
\eeq
The EA is related to the coincidence limits
\beq
\lim\limits_{x \to x^\prime}{\hat a}_k(x,x^\prime)
\,=\,{\hat a}_k\big| \,.
\label{SchDW-6}
\eeq
For the general operator (\ref{O}), the linear
term can be absorbed into the covariant derivative
$\,\na_\mu\to{\cal D}_\mu=\na_\mu+{\hat h}_\mu$,
with the following commutator:
\beq
{\hat S}_{\mu\nu}\,=\,{\hat 1}[\na_{\mu},\na_{\nu}]
-(\na_\nu{\hat h}_\mu-\na_\mu{\hat h}_\nu)
-[{\hat h}_\nu ,{\hat h}_\mu ]\,.
\label{SchDW-7}
\eeq
In this way we arrive at the well-known formulas
\beq
{\hat a}_1\big|\,=\,{\hat a}_1(x,x)\,=\,{\hat P}
\,=\,{\hat \Pi}
+ \frac{\hat 1}{6}\,R - \na_\mu{\hat h}^\mu
- {\hat h}_\mu{\hat h}^\mu
\label{SchDW-8}
\eeq
and
\beq
{\hat a}_2\big|\,=\,{\hat a}_2(x,x)\,=\,
\frac{{\hat 1}}{180}\,(R_{\mu\nu\al\be}^2-R_{\al\be}^2+{\Box}R)
\,+\,\frac12\,{\hat P}^2\,+\,\frac16\,({\Box}{\hat P})
\,+\,\frac{1}{12}\,{\hat S}_{\mu\nu}^2\,.
\label{SchDW-9}
\eeq
One can derive the next coefficients ${\hat a}_3\big|$  and
${\hat a}_4\big|$ \cite{Gilkey,avram-tes}, but we do not
present these (more bulky) expressions here.

The coefficients ${\hat a}_k\big|$
enable one to analyze the EA in a given space-time
dimension for numerous field theory models. For instance, in the
two-dimensional space-time ${\hat a}_1\big|$  defines logarithmic
divergences. In four-dimensional space-time ${\hat a}_2\big|$
defines logarithmic divergences, while ${\hat a}_1\big|$ defines
quadratic divergences. In six-dimensional space-time ${\hat a}_3\big|$
defines logarithmic divergences while  \ ${\hat a}_2\big|$ \
defines quadratic divergences and ${\hat a}_1\big|$ defines
quartic divergences.

An important observation is that the general expressions for the
coefficients ${\hat a}_k\big|$ do not depend on the space-time
dimension \cite{BrowCassidy}. However, the particular traces for a given theory
do have such dependence.
As we have already mentioned in the Introduction, the logarithmic
divergences are universal and scheme-independent. Then, as far
as the coincidence limits ${\hat a}_k\Big|$ are universal in
the space-time dimension where the given coefficient defines
logarithmic divergences, they can be non-universal in other
dimensions. It is easy to see what this means. The finite part
of EA in $d=4$ is given by a sum of all ${\hat a}_k\big|$ with
$k>2$. As far as these coefficients are scheme-dependent in $d=4$,
we can expect that the finite part of EA will be non-universal,
for example the (\ref{com}) may be not satisfied.

From the arguments presented above we can figure out how to
verify the presence of MA in the general fermionic determinant
(\ref{rel}). One has to derive the difference (\ref{com})
between the traces ${\hat a}_k\big|$ for the operators in
an arbitrary dimension $n$. The expected result is that such
a difference vanish for ${\hat a}_1\big|$ in (and only in)
the case of $n=2$, for ${\hat a}_2\big|$ only in the case of
$n=4$, for ${\hat a}_3\big|$ only in the case of $n=6$, etc.

This program has been realized in \cite{QED-form} for the
particular case of QED in curved space and we meet a perfect
correspondence between general arguments and the output of
direct calculations.

In fact, there is no need to perform cumbersome analysis
of ${\hat a}_3\big|$, because one can directly work with the
particular sum of the Schwinger-DeWitt series. The corresponding
heat-kernel solution has been obtained independently by
Barvinsky and Vilkovisky \cite{bavi90} and Avramidi
\cite{Avramidi89}, and it was used in \cite{apco}
for calculating the complete form factors and $\beta$-functions
for massive fields.\footnote{Equivalent form factors were in fact
calculated earlier for the theory with non-zero temperature
in \cite{GusZel}, see also \cite{Milton} for qualitatively
similar expressions in QED.} So, we can safely restrict ourselves
by considering ${\hat a}_1\big|$, ${\hat a}_2\big|$ and the
form factors.

\section{Particular cases of MA}

In the general case of fermionic operator (\ref{act}) with 
conjugate operators (\ref{H1}) and (\ref{opt 2}), one can take 
care of the most simple coefficient of ${\hat a}_1\big|$ to 
arrive at the criteria of existence for the MA. The calculations 
of the traces can be done by using Eq. (\ref{SchDW-8}) and the 
results are as follows:
\beq
\nonumber
a_1^{(1)}(n, \hat{\phi})
&=& \frac{1}{2} \int d^nx
\sqrt{- g}\,\Big\{2(n-1)m \tr(\hat{\phi})
+\frac{(n-2)}{2}\tr(\hat{\phi}\hat{\phi})
+\frac{1}{2}\tr(\hat{\phi}\ga^{\mu}\hat{\phi}\ga_{\mu})\Big\}\,,
\\
a_1^{(2)}(n, \hat{\phi})
&=& \frac{1}{2} \int d^nx
\sqrt{- g}\,\Big\{2(n-1)m \tr(\hat{\phi})+i
\tr(\na_{\mu}\hat{\phi}\ga^{\mu})+\frac{1}{2}\tr(\hat{\phi}
\ga^{\mu}\hat{\phi}\ga_{\mu})\Big\}\,.
\label{29}
\eeq
The difference between these two expressions can be presented as
\beq
a_1^{(1)}(n, \hat{\phi})-a_1^{(2)}(n, \hat{\phi})
\,=\, \frac{1}{4}\int d^nx \sqrt{- g}\,\De_1
\nonumber
\\
\De_1\,=\,\Big\{(n-2)\tr(\hat{\phi}\hat{\phi})
-2i \tr(\na_{\mu}\hat{\phi}\ga^{\mu})\Big\}\,.
\label{30}
\eeq
We can see that this difference consists of two terms. The first
one is proportional to $n-2$, exactly as we have anticipated in
the previous section from general qualitative arguments. According 
to what we have discussed, this term does vanish in the dimension 
$n=2$, where it defines the logarithmic UV divergence. However, 
due to the $n-2$ factor, it does not vanish in $n\neq 2$, and 
hence the quadratic divergence in $n=4$ is scheme-dependent. 
Another part of (\ref{30}) is the surface term, which is also 
quite remarkable, but for different reason. First of all, this 
kind of ambiguity is not related to the mass of quantum field 
and therefore has absolutely different origin compared to the 
terms of the first type. As it was discussed previously in the 
literature on conformal anomaly \cite{duff94,anom-03}, the total 
derivative in the counterterms, in the classically conformal 
massless theories, contributes to the local terms in the 
anomaly-induced EA. As a consequence, these local terms have 
much greater degree of ambiguity than the non-local terms in the 
anomaly-induced EA which can be classified in a regular way 
\cite{Deser93}. Finally, the difference (\ref{30}) includes two 
terms of very different origin which represent two distinct 
types of the QFT ambiguities and hence can not cancel.

Despite it is technically possible to perform the analysis at 
higher orders and obtain the expressions similar to (\ref{30})
for higher Schwinger-DeWitt coefficients, these expressions
are very cumbersome and their sense is sometimes unclear. For 
this reason, it is better to consider only the most interesting 
terms in (\ref{gen}) and do it separately. Let us derive the 
relation (\ref{30}) for a few particular cases.

\vskip 2mm
\noindent
{\Large $\bullet$} \ \ {\large\it Yukawa theory.} \ \
We have $\hat{\phi}=h \phi \hat{1}$. Then
\beq
\De_1\,=\,n(n-2)h^{2}\phi^{2}
- 2 i h \na_{\mu}\phi \tr(\ga^{\mu})=\,n(n-2)h^{2}\phi^{2}\,,
\label{32}
\eeq
\vskip 2mm

\noindent
{\Large $\bullet$} \ \ {\large\it QED.} \ \

The operator $\hat{\phi}$ assumes the form
$\hat{\phi}= e A_{\mu}\ga^{\mu}$.
According to Eq. (\ref{30}), we find (see also \cite{QED-form})
\beq
\De_1\,=\,(n-2)e A_{\mu}A^{\mu}\,-\,2 i \na_{\mu}A^{\mu}\,.
\label{33}
\eeq
\vskip 2mm

\noindent
{\Large $\bullet$} \ \ {\large\it Anomalous magnetic moment.}
\ \
In this case
\ $\hat{\phi}= -\frac{\mu_B}{2}\si_{\mu \nu}F^{\mu \nu}$.
Using Eq.(\ref{30}) we arrive at
\beq
\De_1
&=& (n-2)\dfrac{{{\mu}_{B}}^{2}}{4}F^{\mu \nu}F_{\al \be}
\tr[\si_{\mu \nu}\si^{\al \be}]
\,+\,
i \mu_B \na_\mu F^{\al \be} \tr \big[\si_{\al \be}\ga^{\mu}\,]
\nonumber
\\
&=& n(n-2)\dfrac{{{\mu}_{B}}^{2}}{2}F^{\mu \nu}F_{\mu \nu}.
\label{34}
\eeq
\vskip 2mm

\noindent
{\Large $\bullet$} \ \ {\large\it Torsion.} \ \
In case of absolutely antisymmetric torsion we have
\ $\hat{\phi}= \eta \ga^5\ga^\mu S_\mu$.
\\
As far as $\ga^5$ is defined only in $n=4$, we consider only
this particular dimension. Replacing operator
$\hat{\phi}= \eta \ga^{5}\ga^{\mu} S_{\mu}$ into Eq.(\ref{30})
we arrive at
\beq
\De_1 &=&
2\eta^{2}S_{\mu} S_{\nu}
\tr\big[\ga^{5}\ga^{\mu}\ga^{5}\ga^{\nu}\big]
\,-\, 2 i \eta (\na_{\mu} S_{\nu})
\tr \big[\ga^{5}\ga^{\nu}\ga^{\mu}\big]
\,=\, -8\,\eta^2 S_{\mu}S^{\mu}\,.
\label{39}
\eeq
One can see that in this case there is only one type of
anomalous terms.

\section{Full calculation for Yukawa model}
\label{S1}

Let us now perform complete analysis for a simplest case
of Yukawa model which we have already mentioned in (\ref{32}).

\subsection{Second Schwinger-DeWitt coefficient}
The calculation of the second Schwinger-DeWitt coefficient
can be done in a usual way and provides the following result
for the two calculational schemes (\ref{H1}) and (\ref{opt 2})
in $n$ space-time dimensions:
\beq
a_2^{(k)}(n)\big| \,=\, \int d^nx\sqrt{-g}\left\{
A_k\phi + B_k\phi^2 + C_k\phi^3
+ D_k\phi^4 + E_k\right\}\,,\quad k = 1,2
\label{8}
\eeq
with
\beq
A_1 &=& \frac{nmh}{12}(3-n)R + \frac{nm^3h}{3}(n-3)(n-1)
+ \frac{nh^2}{6}(n-1)\Box\phi\,,
\nonumber
\\
A_2 &=& \frac{nmh}{12}(3-n)R + \frac{nm^3h}{3}(n-3)(n-1) +
\frac{n^2h^2}{12}\Box\phi\,,
\nonumber
\\
B_1 &=& \frac{nh^2}{24}(3-n)R + \frac{nm^2h^2}{6}(9-4n+2n^2)
\,,\;\;\;\;
B_2 = \frac{nh^2}{48}(2-n)R + \frac{nm^2h^2}{4}(n-2)(n-1)\,,
\nonumber
\\
C_1 &=& \frac{nmh^3}{3}(n-3)(n-1)\,,\;\;\;\;
C_2 = \frac{nmh^3}{12}(n-1)\,,
\nonumber
\\
D_1 &=& \frac{nh^4}{12}(n-3)(n-1)\,,\;\;\;\;
D_2 = \frac{n^2h^4}{96}(n+2)\,,
\nonumber
\\
E_1 &=& \frac{nh^2}{12}(n-1)(\nabla\phi)^2
    - \frac{nmh}{6}\Box\phi\,,\qquad
E_2 = \frac{nh^2}{24}(n-2)(\nabla\phi)^2
    - \frac{nmh}{6}\Box\phi\,.
\label{ABCDE}
\eeq

The difference $a_2^{(1)}(n) - a_2^{(2)}(n)$ can be written in
the form
\beq
\nonumber
a_2^{(1)}(n)\big| &-& a_2^{(2)}(n)\big| \,=\,
\int d^nx\sqrt{-g}\,\Big\{
\frac{1}{4}\,\Big[m^{2}h^{2}\phi^{2}+mh^3\phi^3
+ \frac{7}{24}h^{4}\phi^{4}\Big](n-4)^3
\nonumber
\\
&+&
\frac14\, \Big[
 7 m^2h^2\phi^2
- \frac{1}{12}Rh^2\phi^2
- \frac{1}{6}(\na \phi)^2h^2
+7 mh^3\phi^3
+ \frac{25}{12}h^4\phi^4\Big](n-4)^2
\nonumber
\\
&+&
\Big[
 3m^{2}h^{2}\phi^{2}
- \frac{1}{12}Rh^{2}\phi^{2}
- \frac{1}{6}(\nabla \phi)^{2}h^{2}
+3mh^{3}\phi^{3}
+ \frac{11}{12}h^{4}\phi^{4}\Big] (n-4)
\\
&+&\frac{1}{3}h^{2}\Box{\phi}^{2}
\Big\}.
\nonumber
\label{diff a2}
\eeq
It is easy to see that the difference consists of two kinds of
terms. All but the last term do vanish in and only in the four
dimensional case, exactly as the difference in the
$a_1^{(1)}(n) - a_1^{(2)}(n)$ vanish in two dimensions. Obviously,
the structure of these terms confirm our consideratio about the 
universality of the dynamical terms in logarithmic UV divergences 
and, at the same time, the non-universality of power-like 
divergences and finite terms in the case $n\neq 4$. The last term 
in Eq. (\ref{diff a2}) has absolutely different origin. It shows 
the non-universality of surface terms in the logarithmic UV 
divergences. As we already know from the second article in 
\cite{anom-03}, the ambiguity in the term $\Box{\phi}^2$ in the 
one-loop divergences goes, in the massless case, to the ambiguity 
in the corresponding term in the trace anomaly and finally results 
in the ambiguous finite term $R \phi^2$ in the anomaly-induced 
effective action. All in all, our general arguments are confirmed 
here.

\subsection{Calculation of form factors and $\be$-functions}
\label{S2}

The calculation of form factors has been described in full
details in \cite{QED-form,apco,bexi}, so we shall just give the
result of the calculations in our case. The one-loop
contribution to the EA can be presented in the form
\beq
{\bar \Ga}^{(1)}\,=\,
\frac{1}{2} \int d^4x\sqrt{-g}\,
\Big\{
\na_\mu\phi\,k_{{wipf}}(a)\,\na^\mu \phi
+ \phi^2\, k_{R\phi^2}(a)\, R
+ \phi^2\, k_{\phi^2\phi^2}(a)\, \phi^2 \Big\}\,,
\label{111}
\eeq
where the form factors $k$ are defined in terms of useful notations
\beq
Y = 1-\frac{1}{a}\ln \left(\frac{2+a}{2-a}\right)\,,
\quad
a^2 = \frac{4\Box}{\Box - 4m^{2}}\,.
\label{14}
\eeq
We have found the following two sets of form factors corresponding
to the calculational schemes (\ref{H1}) and (\ref{opt 2}).
\beq
k^{(1)}_{R\phi^2}(a) 
&=& -\frac{h^2}{9(4\pi)^2\,a^2}(-14a^2 + 45Ya^2 - 168Y)\,,
\\ \nonumber
k^{(1)}_{\phi^2\,\phi^2}(a) 
&=& \frac{2h^4}{3(4\pi)^2\,a^2}(-8a^2 + 27Ya^2 - 96Y)\,,
\\ \nonumber
k^{(1)}_{kin}(a) 
&=& -\frac{2h^2}{3(4\pi)^2\,a^2}(a^2 + 12Y)\,,
\label{16}
\eeq
within the first calculational scheme (\ref{H1}) and the form factors
\beq
k^{(2)}_{R\phi^2}(a) 
&=& -\frac{h^2}{3(4\pi)^2\,a^2}(-3a^2 + 10Ya^2 - 36Y)\,,
\\
\nonumber
k^{(2)}_{\phi^2\,\phi^2}(a) 
&=& \frac{2h^4}{3(4\pi)^2\,a^2}(6Ya^2 - a^2 - 12Y)\,,
\\ 
\nonumber
k^{(2)}_{kin}(a) 
&=& -\frac{h^2}{3(4\pi)^2\,a^2}(a^2 + 3Ya^2 + 12Y)\,,
\label{17}
\eeq
associated to the second scheme (\ref{opt 2}). The UV $(a\to 2)$ limits
of the two expressions do coincide,
\beq
\nonumber
\lim_{a \to 2}k^{(1,2)}_{R\phi^2}(a) 
&=& -\dfrac{h^{2}}{6(4\pi)^2} \ln(a-2),
\\
\nonumber
\lim_{a \to 2}k^{(1,2)}_{\phi^2\,\phi^2}(a) 
&=& \frac{h^{4}}{(4\pi)^{2}} \ln(a-2),
\\
\lim_{a \to 2}k^{(1,2)}_{kin}(a) 
&=& -\frac{h^{2}}{(4\pi)^{2}} \ln(a-2).
\label{18}
\eeq
The reason is that these limits are related to the logarithmic
divergences in $n=4$ and are therefore universal. However, this
is not true for the form factors themselves, as can be seen from
the expressions (\ref{16}) and (\ref{17}). In particular, the
IR limit $(a\to 0)$ for the same form factors are different,
\beq
\nonumber
\lim_{a \to 0}k^{(1)}_{R\phi^2}(a)
&=& \dfrac{11h^{2}}{60(4\pi)^{2}}a^{2} + {O}({a}^{4}),
\\
\nonumber
\lim_{a \to 0}k^{(2)}_{R\phi^2}(a)
&=& \dfrac{23h^{2}}{180(4\pi)^{2}}a^{2} + {O}({a}^{4}),
\\
\nonumber
\lim_{a \to 0}k^{(1)}_{\phi^2\,\phi^2}(a)
&=& -\dfrac{7h^{4}}{10(4\pi)^{2}}a^{2} +
{O}({a}^{4}),
\\
\nonumber
\lim_{a \to 0}k^{(2)}_{\phi^2\,\phi^2}(a)
&=& -\dfrac{7h^{4}}{30(4\pi)^{2}}a^{2} +
{O}({a}^{4}), 
\\ 
\nonumber
\lim_{a \to 0}k^{(1)}_{kin}(a)
&=& \dfrac{h^{2}}{10(4\pi)^{2}}a^{2} +
{O}({a}^{4}), 
\\
\lim_{a \to 0}k^{(2)}_{kin}(a)
&=& \dfrac{2h^{2}}{15(4\pi)^{2}}a^{2} +
{O}({a}^{4}).
\label{19}
\eeq

Another way to observe the MA in massive theories is though
the physical $\be$-functions. Such $\be$-functions for the
effective charge $C$ can be defined in the framework of the
momentum-subtraction renormalization scheme as
\beq
\be_{C}=\lim_{n \to 4}M \frac{dC}{d M}\,,
\label{192}
\eeq
where the subtraction of divergences is performed at
$p^{2} = M^{2}$, $M$ being the renormalization point.
This is indeed different from the Minimal Substraction
scheme $\beta$-function for the same quantity, which is
given by
\beq
\be_{C}^{MS}\,=\,\lim_{n \to 4}\mu \frac{dC}{d\mu}\,.
\label{191}
\eeq
Both statements apply also to the $\ga$-functions $\ga_{kin}$,
which are related to the renormalization of the kinetic terms
in the scalar field action. The derivative (\ref{192}) can be
expressed in terms of parameter $a$ as
\beq
-\, p \,\frac{dC}{d p}\,=\,(4-a^2)\,\frac{a}{4}\,\frac{dC}{d a}
\label{20}
\eeq
of the form factors in the polarization operator.
Using this procedure, we arrive at the following UV and IR
limits of the corresponding $\be$-functions
\beq
\nonumber
{\be}^{(1)}_{\xi}
&=&\frac{h^{2}}{12(4\pi)^2\,a^2}
\Big\{a^{2}(15a^{2}-56)+(228a^{2}-672-15a^{4})Y\Big\},
\\ \nonumber
{\be}^{(2)}_{\xi}
&=&\frac{h^{2}}{6(4\pi)^2\,a^2}
\Big\{a^{2}(5a^{2}-18)+(74a^{2}-216-5a^{4})Y\Big\},
\\ \nonumber
{\be}^{(1)}_{\la}
&=&-\frac{h^{4}}{2(4\pi)^2\,a^2}
\Big\{a^{2}(9a^{2}-32)+(132a^{2}-384-9a^{4})Y\Big\},
\\ \nonumber
{\be}^{(2)}_{\la}
&=&-\frac{h^{4}}{(4\pi)^2\,a^2}
\Big\{a^{2}(a^{2}-2)+(10a^{2}-24-a^{4})Y\Big\},
\\ \nonumber
\ga^{(1)}_{kin}
&=&\frac{2h^{2}}{(4\pi)^2\,a^2}
\Big\{a^{2}+(12-3a^{2})Y\Big\},
\\
\ga^{(2)}_{kin}
&=&\frac{h^{2}}{4(4\pi)^2\,a^2}
\Big\{a^{2}(a^{2}+4)+(48-8a^{2}-a^{4})Y\Big\}.
\label{beta}
\eeq

The UV limit $a\to 2$ in the complete $\be$-functions (\ref{beta})
correspond to the simple $MS$-scheme expressions and is the same
for the two calculational approaches,
\beq
{\be}^{(1,2) \;UV}_{\xi}
&=& -\,p\frac{dk^{(1,2)}_{R\phi^2}(a)}{dp}
\,=\,  \frac{h^2}{3(4\pi)^2}\,,
\nonumber
\\
{\be}^{(1,2) \;UV}_{\la}
&=& -\,p\frac{dk^{(1,2)}_{\phi^2\,\phi^2}(a)}{dp}
\,=\, - \frac{2h^4}{(4\pi)^2}\,,
\nonumber
\\
\ga^{(1,2) \;UV}_{kin}
&=& -\,p\frac{dk^{(1,2)}_{\na\phi\,\na\phi}(a)}{dp}
\,=\,  \frac{2h^2}{(4\pi)^2}\,,
\label{21}
\eeq
In the opposite, IR, limit the situation is quite different,
indicating an ambiguity in the Appelquist and Carazzone theorem,
\beq
\nonumber
{\be}^{(1) \;IR}_{\xi}=-p\frac{dk^{(1)}_{R\phi^2}(a)}{dp}
\,=\, \frac{11h^2}{30(4\pi)^{2}}a^2 + O(a^4)\,,
\\
\nonumber
{\be}^{(2) \;IR}_{\xi}=-p\frac{dk^{(2)}_{R\phi^2}(a)}{dp}
\,=\, \frac{23h^2}{90(4\pi)^{2}}a^2 + O(a^4)\,,\\ \nonumber
{\be}^{(1) \;IR}_{\la}=-p\frac{dk^{(1)}_{\phi^2\,\phi^2}(a)}{dp}
\,=\, -\frac{7h^4}{5(4\pi)^{2}}a^2 + O(a^4)\,,\\ \nonumber
{\be}^{(2) \;IR}_{\la}=-p\frac{dk^{(2)}_{\phi^2\,\phi^2}(a)}{dp}
\,=\, -\frac{7h^4}{15(4\pi)^{2}}a^2 + O(a^4)\,,\\ \nonumber
{\ga}^{(1) \;IR}_{kin}=-p\frac{dk^{(1)}_{\na\phi\,\na\phi}(a)}{dp}
\,=\, \frac{h^2}{5(4\pi)^{2}}a^2 + O(a^4)\,,\\
{\ga}^{(2) \;IR}_{kin}=-p\frac{dk^{(2)}_{\na\phi\,\na\phi}(a)}{dp}
\,=\, \frac{4h^2}{15(4\pi)^{2}}a^2 + O(a^4)\,.
\label{22}
\eeq
In the space with Euclidean signature we have, for $p^2 \ll m^2$,
the relation $a^2 \propto p^2/m^2$ in the low-energy IR limit.
Then we can see that in all cases the decoupling in Eqs. (\ref{22})
is quadratic, according to the Appelquist and Carazzone theorem,
but the coefficients depend on the choice of calculational scheme, 
that is whether we use operator $\hat{H}^{*}_1$ from (\ref{H1}) or
operator  $\hat{H}^{*}_2$ from (\ref{opt 2}). Let us note that
from physical viewpoint the first choice with $\hat{H}^{*}_1$
is much better because it helps to preserve the gauge invariance
in QED \cite{QED-form}. This means we have to make such a choice
in the {\it ad hoc} manner. Definitely, it is important to be aware
of the possible risks of making an alternative choice.

\section{Conclusions and discussions}
\label{Con}

We have explored in details an ambiguity which takes place
in the derivation of fermionic functional determinants by
means of the heat kernel method. There are two kind of
ambiguities, which have essentially different origins. The
first one takes place only in case of massive theories and
shows the deep importance and universality of the logarithmic
UV divergences. The divergences can be always removed by
renormalization procedure, but its remnants in form of
leading logarithmic behavior of form factors do remain and
represent the most stable part of quantum corrections.
An important observation, from our viewpoint, is that
the universality of logarithmic UV divergences should hold
in {\it any} spacetime dimension. As a consequence of this
feature, the finite part of one-loop EA in massive theories
becomes scheme-dependent. In case of fermionic determinants
this can be seen in the form of non-local multiplicative
anomaly. It is clear that this kind of ambiguity can not
be seen in massless theories, because in this case the EA
is much more controlled by leading logarithmic terms. It
would be very interesting to find another examples of such
an ambiguity for other (non-fermionic) theories, and we
hope to find such examples in future.

Another sort of ambiguity does not depend on whether the quantum
field is massive or massless, it occurs in the divergent
total-derivative terms. These terms do not depend on whether 
the theory is massive or massless. For the particular case of
fermionic determinants this means independence on whether the 
initial classical theory is conformal or not. And in case of 
conformal theories, these divergent surface terms are known to 
contribute to the conformal
anomaly and finally to the local terms in the anomaly-induced
EA, where they can be removed by adding finite local counterterm.
Therefore, this ambiguity is quite different from the one of the
first kind, which is essentially non-local and takes place only
in massive theories. The common point is that both of them {\it can
not} be compensated by the change of coefficients in the infinite
local counterterms, which are introduced in the process of
renormalization.

\section*{Acknowledgments}
Authors are thankful to I.V. Tyutin and S. Ranjbar-Daemi
for useful advises which led them to Ref. \cite{Salam-51}.
D.P. is grateful to FAPEMIG for supporting his PhD project.
G.B.P. and I.Sh. are grateful to CNPq, FAPEMIG and CAPES
for partial support.

\renewcommand{\baselinestretch}{0.9}

\end{document}